# Fermi surface and superconducting gap of triple-layered $Bi_2Sr_2Ca_2Cu_3O_{10+\delta}$

[a)]R. Müller, [a)]C. Janowitz, [a)]M. Schneider, [a)]R.-St. Unger, [a)]A. Krapf, [a)]H. Dwelk, [a)]A. Müller, [a)]L. Dudy, [a)]R. Manzke, [b)]H. Höchst

[a)] *Institut für Physik, Humboldt-Universität zu Berlin, Invalidenstraße 110, 10115 Berlin, Germany*

[b)] *Synchrotron Radiation Center, Stoughton, WI, USA*

**Abstract**

We present a comprehensive study performed with high-resolution angle-resolved photoemission spectroscopy on triple-layered $Bi_2Sr_2Ca_2Cu_3O_{10+\delta}$ single crystals. By measurements above $T_C$ the Fermi surface topology defined by the Fermi level crossings of the $CuO_2$-derived band was determined. A hole-like Fermi surface as for single and double-$CuO_2$ layered Bi-based cuprates is found, giving new input to the current debate of the general Fermi surface topology of the high $T_C$ superconductors. Furthermore, we present measurements of the superconducting gap of Bi-2223 and show that there are clear indications for a strong anisotropy of the superconducting gap. The universal properties of this phase in comparison to the other Bi-based cuprates will be discussed.

The capability of angle-resolved photoemission spectroscopy (ARPES) to probe the energy and momentum relations directly makes it a powerful tool to study the electronic structure of high temperature superconductors (HTSC) and other strongly correlated systems. Impressive progress has been achieved over the last decade by ARPES in order to resolve the characteristics of high-$T_C$ superconductors. Major discoveries have been made on both the normal and the superconducting state. As examples may be mentioned the hole-like Fermi surface due to the weakly dispersing $CuO_2$ derived band, the anisotropy of the superconducting gap and observations of a pseudo gap [1]. Up to now, most of these studies were done on Bi-2212 single crystals with two $CuO_2$-layers per unit cell and to a much smaller extent on Bi-2201 with one such layer [1].

A central issue in the physics of high-$T_C$ superconductors is the role of coupling between the two-dimensional copper–oxygen planes in producing superconductivity. Since in the layered compounds of type $A_2B_2Ca_{(n-1)}Cu_nO_{2n-4+\delta}$, where A is either Bi, Hg or Tl and B is Sr for the Bi-systems and Ba for the Hg- and Tl-compounds, the superconducting transition temperature $T_c$ increases with increasing number n of square $CuO_2$ layers per unit cell up to n=3. It is therefore of great interest whether the normal state properties and also the symmetry of the superconducting state order parameter will be the same in all systems where the coupling between the layers can be expected to vary considerably. This is due to a strong variation of the distance between neighbouring $CuO_2$ planes which is 11.6 Å for the one-layer compound and about 3.2 Å for higher stages. Recently for n=2 Bi-2212 a bi-layer splitting could be detected [2]. This could be taken as a strong indication for distinct coupling of the $CuO_2$ planes along the c-axis in Bi-based cuprates. Therefore, further studies of the effects on the electronic structure are necessary and are crucial for an understanding of high temperature superconductivity. Interestingly, $T_C$ decreases for n = 4 and higher stages. For n ≥ 3 the $CuO_2$ planes are no longer equivalent and an inhomogeneous distribution of the charge carriers among the different types of layers is expected [3,4]. While sufficient data on the momentum, temperature and doping dependence of the excitation gap of the bilayer compound Bi-2212 [1,5-7] exist, only few photoemission results are published on the n=3- phase in single crystalline form [8-10].

In the present work new high-resolution photoemission results on the dispersion of the emission peak at $E_F$ in the normal state and the derived Fermi surface topology of $Bi_2Sr_2Ca_2Cu_3O_{10+\delta}$ single crystals is presented. In addition, spectra below the critical temperature of different points of the Brillouinzone will be given for the first time. The results are discussed in connection to previous measurements of n = 1 and n = 2 Bi-cuprates.

The crystal structure of Bi-2223 is almost identical to the well-known Bi-2212 and Bi-2201 systems, except for three $CuO_2$ planes per unit cell instead of two or one. The growth of Bi-2223 single crystals has been investigated for the first time by Matsubara et al. [11], who proposed a high temperature annealing procedure in order to convert Bi-2212 single crystals or whiskers to Bi-2223 in a so-called lead bath. Although this procedure seem to work satisfactory for whiskers, for crystals the crystal quality as well as the single phase purity came out quite poor. This has been confirmed by our own investigations [12,13]. Instead of this, the Bi-2223 single crystals used in this investigation were obtained by crystal conversion of the not commonly used Bi-2112 phase through high temperature annealing in a powder composition $Bi_2Sr_2Ca_4Cu_6Pb_{0.5}O_{14}$. By optimizing the annealing temperature and the initial Bi-2112 phase the crystal quality has been crucially improved. The details of the growth process will be described elsewhere [14]. The samples were rectangular shaped with the long side along the crystallographic a-axis, as confirmed by Laue-diffraction and *in situ* low energy LEED, and have a typical size of (0.5×0.8) mm$^2$. Characterization of the crystals was performed by energy dispersive EDX and AC-susceptibility measurements. The composition of our samples is typically $Bi_{2.23}Pb_{0.02}Sr_{1.73}Ca_{2.12}Cu_{2.92}O_{10+\delta}$. It is important to note that lead is not or at least to a very small amount incorporated in the crystals. The superconducting transition of the magnetic suszeptibility occurs in one step, indicating a full conversion of the Bi-2112 phase to the Bi-2223 phase. $T_c$ is 108 K at a $\Delta T_C$ of 3-4 K, indicating probably a slight underdoping of the samples.

The ARPES spectra taken with hν = 21 eV photon energy above $T_C$ were recorded with synchrotron radiation at BESSY, Berlin, at the U125-1 PGM beamline with the high-resolution photoemission station HIRE-PES [15]. The temperature of the sample was T=120 K, the overall energy resolution was 35 meV and the angle resolution 1°. The sample was mounted with the Cu-O bonds in the horizontal plane (ΓM or (0,0) → (π,0)) and the emitted photoelectrons were collected in the respective high symmetry directions, i.e. in that plane and out of it. Accordingly for in plane emission the polarization was even and for out of plane emission the polarization was mixed. The ARPES-spectra shown for hν=22 eV and 20K were performed at the Synchrotron Radiation Center SRC, Madison, using a plane grating monochromator at an undulator beam line with a resolving power of $10^4$ at $10^{12}$ photons/s combined with a SCIENTA-200 electron analyzer in an angle resolving mode. A typical measurement was performed with an angular resolution window of 0.2° and an energy resolution of 16 meV (FWHM). The samples were mounted with the Cu-O bonds in the horizontal plane and the emitted photoelectrons were collected within the polarization plane. This gave even polarization geometry for emission along ΓM. The Fermi level was determined by measuring the Fermi edge of a Au film in electrical contact with the sample.

Several ARPES cuts along high symmetry lines of $Bi_2Sr_2Ca_2Cu_3O_{10+\delta}$, taken in the normal conducting state at T=120 K are shown in FIG. 1 (left panels). Information on the Fermi surface is obtained from the points at which the bands cross the Fermi energy. These spectra of k=$k_F$ are marked by a thick line in FIG. 1. These k-values were taken were the midpoint of the leading edge has the largest kinetic energy. In the right panel of FIG. 1 these crossing points (open circles) are depicted together with a Fermi surface as given for n=2 Bi-2212 by Ding et al. [16]. Additionally, the crossings points from our previous photoemission measurement along ΓY [8] were also incorporated (filled circles). The thick curves in the right panel stem from the Fermi surface of the main $CuO_2$ derived band (main FS). The thin curves result from an umklapp of the main band by the reciprocal superlattice vector of **q**=(0.21π,0.21π) associated with the reconstruction of the BiO planes in ΓY direction. One observes clear Fermi surface crossings due to main $CuO_2$ band along ΓX (and ΓY) and two additional along XMY. Besides the main Fermi surface, along the ΓY direction distinct crossings due to the diffraction replica have been found [8]. The EDC series along ΓM reveals near M the dispersion of a flat band, also reported for n=1 and n=2 material [17,18] leading to an extended saddle point located close to the Fermi level. A Fermi surface crossing is observable on this line only at low angles, attributed to the crossing of the diffraction replica band. All these observations are in striking correspondence with the photoemission results of n=1 and n=2 Bi- cuprates. In particular, the two crossing points due to the main band along XMY must be interpreted as strong hints for a hole-like topology of the Fermi surface also for n=3 Bi-2223.

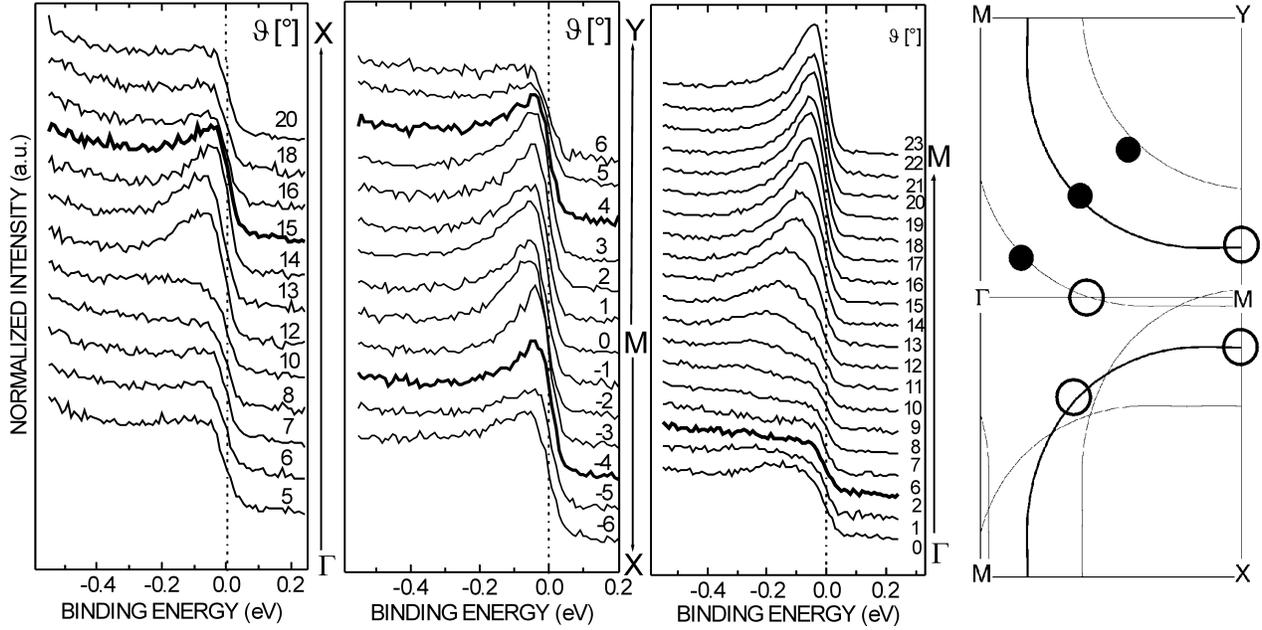

Figure 1. Left panels: High-resolution ARPES series of $Bi_2Sr_2Ca_2Cu_3O_{10+\delta}$ in the normal conducting state (T=120 K). Shown are the dispersion directions ΓX, XMY and ΓM with 21 eV photon energy. Right panel: Comparison of the experimentally found crossing points through $E_F$ (open circles: hν=21eV) and the hole-like Fermi surface proposed for Bi-2212 by Ding et al. [16]. Supplemented are also results from our previous work [8] (filled circles: hν=18eV)

With the knowledge of the shape of the Fermi surface we can focus now on the measurements below $T_C$. For a comparison we have chosen the most interesting k-points, i.e. the M-point, a point on the main FS along ΓY, and one near M on the main FS. For n=2 Bi-2212 the Fermi surface remains unchanged below $T_C$ along the nodal direction of the superconducting gap, which is the ΓX(Y) direction. It is destroyed by the appearance of the gap with its maximal value near the M-point [5].

As for the normal conducting state of Bi-2223, photoemission measurements below $T_C$ are also very rare. Gu et al. [19] have reported on measurements of highly textured $Bi_{1.8}Pb_{0.4}Sr_2Ca_2Cu_3O_{10+\delta}$ and determined a superconducting gap of Δ=29 meV. This is somewhat larger than for the n=2-phase (typically slightly above 20 meV). They found indications of an anisotropic gap. Ponomarev et al. [9] have compared the size of the superconducting gap as determined by tunnelling and by highly resolved photoemission on single crystals and found accordingly a somewhat larger value of 36 meV by both methods. These authors also compiled the gap values of different families of high temperature superconductors to determine general trends.

Our measurements of single crystalline n=3-samples at hν=22eV below $T_C$ are shown in FIG. 2. Results of three points of the Brillouin zone are compared as indicated in the inset. Especially at the M(π,0)-point a very sharp peak with its maximum at 60 meV binding energy occurs followed by a broader and less intense structure at higher binding energy. The ratio of the intensity at the maximum of the peak to the intensity at 500 meV (background) is 3.9. The size of the gap is simply obtained as the energy position of the half height of the leading edge with respect to the Fermi level. Near the M-point a leading edge shift of 37 meV related to the superconducting gap size $\Delta$ is observed, by far larger than commonly observed for n=2-material of approximately 20…24 meV [4,5,20]. The gap closes for a k-vektor along the $\Gamma Y$-line on the Fermi surface (point $F_1$) and has a medium value between these two k-points (eg. point $F_2$). Although these three data points do not permit to give a complete fit to a d-wave gap functional form (because the measured points are not dense enough), they show clearly a very strong anisotropy which could be described by a d-wave gap. Taking our maximal gap value of $\Delta$ = 37 meV for n = 3 together with the corresponding values for n = 1 and n = 2 of 10 meV [21] and 22 meV [4,5], respectively, we derive a linear dependence between $\Delta$ and n. This agrees very well with our previous results and the comparison to tunnelling spectroscopy [9,22]. In contrast to this, $T_C$ does not scale linearly with n [3,4,23].

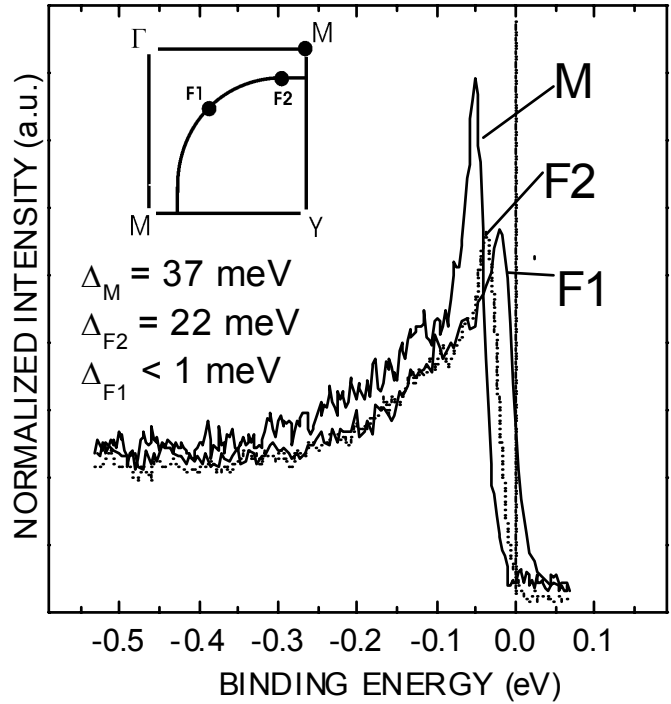

Figure 2. Spectra of triple-layered Bi-2223 taken in the superconducting state at 20 K at different points of the Brillouin zone as indicated in the inset. The photon energy was 22eV. Also given are the experimentally derived values of the superconducting gap.

Taking all measured Fermi level crossings and the flat band behavior along $\Gamma M$ into account, a Fermi surface similar to that found for n=1 Bi-2201 and n=2 Bi-2212 is also valid for the n=3 Bi-2223 samples. With respect to the current Fermi surface controversy, it can be argued that the hole-like Fermi surface centered around X(Y) for ARPES data taken with "low" photon energies (h$\nu \leq$ 22eV) is a universal characteristic of all Bi-cuprates. Furthermore, the Fermi surface for n=1-3 Bi-based cuprates encloses similar areas in k-space.

To discuss the intensity of the emission maximum at $E_F$ (the occupied part of the Zhang-Rice (ZR) singlet band) is difficult because of the strong dependence of photoemission spectra from crystal quality, surface quality, doping and matrix element effects [24,25]. Due to these uncertainties it is not evident, whether the intensity of the peak in the superconducting phase can be compared between the phases and also whether a scaling of the intensity versus number of $CuO_2$ layers n or versus $T_C$ or any other parameter is possible. Instead the unoccupied part can be probed without great uncertainity: X-ray absorption spectroscopy (XAS) probes the bulk properties due to its sampling depth of about 100 nm, the polarization dependence can be explicitly studied and also the effects of self - absorption can be incorporated [26].

In a previous study on opimally doped single-crystalline single-, double-, and triple-layered Bi-cuprates with x-ray absorption it was shown by Manzke et al. [23] that the O 1s pre-peak intensity, which is probably due to the density of the unfilled part of the Zhang-Rice singlet states, indeed scales linearly with the number of $CuO_2$-planes per unit cell n. An interesting ansatz to describe the charge distribution in the layered cuprates was given by Di Stasio et al. [3]. They proposed that the doping in every stage n of the Bi cuprates is performed by an equal charge reservoir, the two Bi-O planes per unit cell with the extra charge density -δ. This induces a density of holes that must be shared by the Cu-O layers, which in principle is easy for the stages n=1 and 2. For pure n=2 samples, the doping of the Cu-O planes is almost at its optimal value resulting in a $T_C$ of 90 K. Here the two Cu-O planes share a charge of +δ. If one assumes that the doping strength from the Bi-O reservoir may be about the same for all stages, then pure n=1 material samples should be extremely overdoped. This is, in fact, observed. However, this is progressively more difficult in triple-layer stages in which the induced density now must be shared nonhomogeneously along the three Cu-O layers. Such a charge distribution can be described by the sheet-charge model proposed by Di Stasio et al. [3]. Now, if the intensity of the pre-peak is due to the number of holes in the Cu-O planes, this would imply for triple-layer material that the δ-value must have been increased by about 50% relative to double-layer material. However, as the superconducting gap value has the same linear dependence on n [9,23] like the pre-peak intensity, we would argue that instead of being due to the total hole density in the Cu-O planes, the intensity of the pre-peak in XAS represents only these holes that directly "serve" superconductivity. This is in line with the linear scaling behavior of the gap size versus the number n of $CuO_2$-planes per unit cell. The linear Δ(n) behavior is not only confirmed for BSCCO by tunneling spectroscopy and ARPES [4,9,22], but also for HBCCO and TBCCO by tunneling spectroscopy [9] and seems to be a universal scaling behavior in the cuprates.

Another important aspect of our n=3 data is the observed anisotropy of the superconducting gap. For n=2 Bi-2212 the d-wave order parameter in the high $T_C$ superconductors is established [18]. Our measurements verify these findings for the triple layer Bi-2223. It is interesting to note that at the M-point, which is clearly away from the Fermi surface, the superconducting gap is higher than near the Fermi surfac crossing along MY. This is in good agreement with recent measurements on Bi-2212 [27].

In summary, we presented an investigation of the Fermi surface and superconducting gap performed with temperature dependent high-resolution photoemission spectroscopy on triple-layered $Bi_2Sr_2Ca_2Cu_3O_{10+\delta}$ single crystals. Our detailed study reveals strong experimental evidence for a hole-like Fermi surface and an anisotropic superconducting gap, which has its maximal value of Δ = 37 meV at M and vanishes along the nodal direction ΓY. These observations are in close correspondence to n = 1 and n = 2 Bi-cuprates near optimal doping and point to an important universal property of this superconductor family. Beyond this, the amount of the superconducting gap of n=3 together with that of n=1 and n=2 scales linearly with the number of $CuO_2$ – planes per unit cell.

During the preparation of this manuscript two new preprints on n=3 Bi-2223 single crystals came to our attention [28,29] which are currently in the referee process. While the former confirms our view of the shape of the Fermi surface [8,10] and of the gap anisotropy, both publications report on a linear scaling of the superconducting gap with $T_C$, but not with n, an issue which may need further studies.

## Acknowledgement

We would like to thank D. Kaiser for enormous help by the preparation and Dr. S. Rogaschewski, Dr. P. Schäfer, C. Ast and B. Stoye for characterization of the samples. We gratefully acknowledge assistance by the staff of BESSY II, especially that of Dr. R. Follath. This work is supported by the BMBF under contract no. 05 SB8KH1/0. Part of this work is based upon research conducted at the Synchrotron Radiation Center, University of Wisconsin-Madison, which is supported by the NSF under Award No. DMR-0084402. One of the authors (R.Manzke) thanks the Deutsche Forschungsgemeinschaft (DPG) for financial support (MA 2371/1).

## References

[1] see e.g. Z.-X. Shen, D.S. Dessau, Physics Reports **253,** 1 (1995); T. Timusk B. Statt, Rep. Prog. Phys. **62**, 61 (1999); T. Tohyama, S. Maekawa, Supercond. Sci. Technol. **13,** 17 (2000); V.M.Loktev, R.M. Quick, S.G. Shapapov, Physics Reports **349**, 1 (2001), and references therein

[2] D. L. Feng, N. P. Armitage, D. H. Lu, A. Damascelli, J. P. Hu, P. Bogdanov, A. Lanzara, F. Ronning, K. M. Shen, H. Eisaki, C. Kim, Z.-X. Shen, J.-i. Shimoyama, K. Kishio Phys. Rev. Lett. **86**, 5550 (2001); Y.-D. Chuang, A. D. Gromko, A. Fedorov, Y. Aiura, K. Oka, Yoichi Ando, H. Eisaki, S. I. Uchida, D. S. Dessau, Phys. Rev. Lett. **87**, 117002 (2001)

[3] M. Di Stasio , K.A. Müller, L. Pietronero, Phys. Rev. Lett. **64**, 2827 (1990)

[4] A. Müller, PhD thesis, Humboldt-Universität Berlin (2000)

[5] M. Randeria, J.C. Campuzano, Proceedings of the International School of Physics, "Enrico Fermi", Varenna, 1997, IOS Press, Amsterdam, 1998, preprint cond-mat/97091077 (1997), and references therein

[6] Z.X. Shen, J.R. Schriefer, Phys. Rev. Lett. **78**, 1771 (1997).

[7] A. G. Loeser, Z.-X. Shen, M. C. Schabel, C. Kim, M. Zhang, A. Kapitulnik, P. Fournier Phys. Rev. B **56**, 14185 (1997).

[8] C. Janowitz, R. Müller, T. Plake, A. Müller, A. Krapf, H. Dwelk, R. Manzke, Physica B **259,** 1134 (1999)

[9] Y.G. Ponomarev, N.Z. Timergalev, K. K. Uk, M.A. Lorenz, G. Müller, H. Piel, H. Schmidt, C.Janowitz, A. Krapf, R. Manzke, Fourth European Conference on Applied Superconductivity, Sitges/Spain, 14.-17.9.99, in: Proceedings of EUCAS, Institute of Physics Conference Series, **167/2**, 245 (2000)

[10] C. Janowitz, R. Müller, M. Schneider, A. Krapf, R. Manzke,  C. Ast, H. Höchst, Physica C **364-365** (2001), in press

[11] I. Matsubara, H. Tanigawa, T. Ogura, H. Yamashita, M. Kinoshita, T. Kawai, Appl. Phys. Lett. **58**, 409 (1991)

[12] A. Krapf, G. Lacayo, G. Kästner, W. Kraak, N. Pruss, H. Thiele, H. Dwelk, R. Hermann, Supercond. Sci. Technol. **4**, 237 (1991)

[13] W. Kraak, A. Krapf, N. Pruss, G. Neubert, S. Rogaschewski, I. Hähnert, W. Wilde, P. Thiele, Phys. Stat. sol. (a) **158**, 183 (1996)

[14] A. Krapf et al., to be published

[15] C. Janowitz, R. Müller, T. Plake, T. Böker, R. Manzke, J. Electron Spectroscopy and Related Phenomena **105,** 43 (1999)

[16] H. Ding, A.F. Bellman, J.C. Campuzano, M. Randeria, M.R. Norman, T. Yokoya, T. Takahashi, H. Katayama-Yoshida, T. Mochiku, K. Kadowaki, G. Jennings, G.P. Brivio, Phys. Rev. Lett. **76**, 1533 (1996)

[17] D. M. King, Z-.X. Shen, D. S. Dessau, D. S. Marshall, C. H. Park, W.E. Spicer, J. L. Peng, Z. Y. Li, R. L. Greene, Phys. Rev. Lett. **73**, 3298 (1994)

[18] D. S. Dessau, Z.–X. Shen, D. M. King, D. S. Marshall, L. W. Lombardo, P. H. Dickinson, A. G. Loeser, J. DiCarlo, C.–H Park, A. Kapitulnik, W. E. Spicer, Phys. Rev. Lett. **71**, 2781 (1993)

[19]   C. Gu, B.W. Veal, R. Liu, A.P. Paulikas, P. Kostic, H. Ding, J.C. Campuzano, B.A. Andrews, R.I.R. Blyth, A.J. Arko, P. Manuel, D.Y. Kaufman, M.T. Lanagan, Phys. Rev. B **51**, 1397 (1995)
[20]   H. Ding, M. R. Norman, J. C. Campuzano, M. Randeria, A. F. Bellman, T. Yokoya, T. Takahashi, T. Mochiku, K. Kadowaki, Phys. Rev. B **54**, R9678 (1996)
[21]   J.M. Harris, P.J.White, Z.-X. Shen, H. Ikeda, R. Yoshizaki, H. Esaki, S. Uchida, W.D. Si, J.W. Xiong, Z.-X. Zhao, D.S. Dessau, Phys. Rev. Lett. **79**, 143 (1997)
[22]   S.I. Vedeneev, A.G.M. Jansen, P. Wyder, Physica B **300**, 38 (2001)
[23]   R. Manzke, R. Müller, M. Schneider, C. Janowitz, A. Krapf, A. Müller, H. Dwelk, J. Superconductivity **13**, 883 (2000)
[24]   S. V. Borisenko, A. A. Kordyuk, S. Legner, C. Dürr, M. Knupfer, M. S. Golden, J. Fink, K. Nenkov, D. Eckert, G. Yang, S. Abell, H. Berger, L. Forró, B. Liang, A. Maljuk, C. T. Lin, B. Keimer, Phys. Rev. B **64**, 094513 (2001)
[25]   R. Manzke, R. Müller, M. Schneider, C. Janowitz, A. Krapf, A. Müller, H. Dwelk, Phys. Rev. B **63**, R100504 (2001)
[26]   W. Frentrup, D. Schröder, R. Manzke, J. Phys. IV France **7**, 509 (1997)
[27]   H. Ding, J.R. Engelbrecht, Z. Wang, J. C. Campuzano, S.-C. Wang, H.-B. Yang, R. Rogan, T. Takahashi, K. Kadowaki, D. G. Hinks, cond-mat/0006143 (2000)
[28]   D. L. Feng, A. Damascelli, K. M. Shen, N. Motoyama, D. H. Lu, H. Eisaki, K. Shimizu, J.-i. Shimoyama, K. Kishio, N. Kaneko, M. Greven, G.D. Gu, X. J. Zhou, C. Kim, F. Ronning, N. P. Armitage, Z.-X. Shen, cond-mat/0108386 (2001)
[29]   T. Sato, H. Matsui, S. Nishina, T. Takahashi, T. Fujii, T. Watanabe, A. Matsuda, cond-mat/0108415 (2001)